\documentclass[runningheads]{llncs}

\usepackage[numbers]{natbib}
\usepackage[utf8]{inputenc}
\usepackage[T1]{fontenc}
\usepackage{amsmath,amssymb,amsfonts}
\usepackage{graphicx}
\usepackage{textcomp}
\usepackage{url}
\usepackage{natbib}

\usepackage{amsmath,amssymb,amsfonts}
\usepackage{textcomp}
\usepackage{xcolor}
\usepackage{url}
\usepackage{algorithm}
\usepackage{algpseudocode}
\usepackage{tikz}
\usetikzlibrary{arrows.meta,positioning,fit,shapes.geometric,backgrounds}
\usepackage{multirow}
\usepackage{graphicx}
\usepackage{booktabs}
\usepackage[utf8]{inputenc}
\usepackage[colorlinks,bookmarksopen,bookmarksnumbered,
                citecolor=red,urlcolor=red]{hyperref}
\usepackage{tabularx}
\usepackage{array}
\usepackage[T1]{fontenc}

\usepackage{moreverb}

\usepackage{booktabs}
\usepackage{multirow}
\usepackage{array}
\usepackage{siunitx}
\usepackage{rotating}
\usepackage{tabularx}
\usepackage{float}
\usepackage{placeins}
\usepackage{algorithm}
\usepackage{algpseudocode}
\usepackage{hyperref}
\usepackage{cleveref}
\usepackage{xcolor}
\usepackage{listings}

\def\tsc#1{\csdef{#1}{\textsc{\lowercase{#1}}\xspace}}
\tsc{WGM}
\tsc{QE}
\begin{document}
\let\WriteBookmarks\relax
\def\floatpagepagefraction{1}
\def\textpagefraction{.001}

\title{Delphi Scanner: efficient and interpretable static malware detection via API sequence modeling}

\author{Bijied Brahimi\inst{1} \and
Vincent Cohadon\inst{1} \and
Gabriel Glazman\inst{1} \and
Rayan Al Mohaize\inst{1} \and
Omran Berjawi\inst{2} \and
Rida Khatoun\inst{2}}

\authorrunning{Brahimi et al.}

\institute{Université Paris Cité, Paris, France
\and
Institut Polytechnique de Paris, Télécom Paris, Palaiseau, France}

\maketitle

\begin{abstract}
Static malware detection for Windows Portable Executable files demands a careful balance between detection effectiveness, computational efficiency, and analytical interpretability. This paper introduces Delphi Scanner, a static malware detection system for Windows PE files that balances efficiency with behavioral interpretation. It uses a convolutional neural network (CNN) to model Windows API sequences to classify PE and a decoupled interpretation layer based on a rule-based layer to categorize APIs into high-level malicious capabilities.  Evaluated on over 190,000 Windows PE files, the system achieves 95.35\% accuracy with a 1.53~MB model footprint. Robustness experiments on 5,647 out-of-distribution MalwareBazaar samples, paired packed and unpacked executables, and three adversarial manipulation strategies confirm generalization beyond the training distribution and resistance to functionality-preserving evasion techniques. Overall, these results demonstrate that API sequence-based static analysis offers a practical, interpretable, and efficient foundation for malware triage in local deployment scenarios.
\end{abstract}


\section{Introduction}
Malicious software targeting the Microsoft Windows ecosystem remains one of the most persistent and impactful threats in contemporary cybersecurity.  Windows executables, distributed in the Portable Executable (PE) format, a standard binary structure for Windows programs and libraries, constitute the primary delivery mechanism through which such malware is distributed. Despite decades of research and the deployment of sophisticated commercial antivirus solutions, malware continues to evolve rapidly in volume and evasiveness, exploiting the widespread adoption of Windows-based systems across personal, corporate, and industrial environments~\cite{avtest2024report}.

This persistent threat landscape has driven sustained interest in automated malware detection techniques.  Traditionally malware detection works in two techniques: Dynamic and static analysis. Dynamic analysis operates by executing suspicious software within an isolated environment and monitoring its runtime behavior, capturing system-level events including API calls, file system and registry modifications, network connections, and memory allocation patterns~\cite{egele2012survey}. This technique can uncover concealed behaviors, but it requires high computational cost and careful environment configuration, and it remains vulnerable to evasion by environment-aware malware. In contrast, static techniques examine executables without execution by inspecting their binary structure, metadata, and embedded artifacts. This approach enables fast, deterministic, and safe analysis, making it attractive for pre-execution screening and large-volume malware triage \cite{ye2017survey}

Early static malware detection relied on manually crafted signatures, which proved brittle against minor code modifications and ineffective against unseen threats. Consequently, research has increasingly adopted machine learning to automatically learn discriminative patterns from static PE file features~\cite{anderson2018ember}, including PE headers, section statistics, and imported libraries~\cite{saxe2015deepmalware}. Among these, Windows API imports are particularly compelling: they offer a compact, semantically meaningful abstraction of program capabilities, revealing dependencies on file system access, network communication, registry manipulation, and process control, and have long served as key indicators for malware analysts~\cite{cesare2014classification}.

Recent deep learning advances have further enabled modeling API usage as sequential data, allowing classifiers to capture local patterns and co-occurrence relationships among API calls~\cite{maniriho2023apimaldetect}. Despite promising detection performance, many existing approaches prioritize accuracy in isolation and overlook practical deployment constraints, including inference latency, memory footprint, explainability, and user trust~\cite{gibert2020rise}. Complex architectures like transformers, while expressive, often introduce substantial computational overhead without commensurate gains for static analysis tasks~\cite{alshomrani2024transformersurvey}. A persistent limitation is interpretability: black-box predictions provide little insight into classification decisions, complicating analyst validation and undermining trust in automated systems~\cite{arp2022and, doshi2017towards}. This highlights a critical gap — current systems struggle to simultaneously balance detection effectiveness, computational efficiency, and meaningful interpretability under local, offline deployment constraints.

Motivated by this challenge, this paper presents Delphi Scanner, a lightweight static malware detection system for Windows PE files that designed to explicitly balance detection effectiveness, inference efficiency, and analytical interpretability within a locally deployed architecture. The system extracts ordered Windows API call sequences from the Import Address Table (IAT) of PE files—including resolution of ordinal-based imports—and encodes them into fixed-length numerical representations that serve as the primary input to a multi-scale one-dimensional convolutional neural network (CNN) classifier optimized for sub-millisecond inference via ONNX Runtime deployment. In parallel, and independently of the classification pipeline, a rule-based behavioral interpretation layer maps the same extracted API sequences to predefined malicious capability categories each aligned with MITRE ATT\&CK technique identifiers~\cite{lipton2018mythos}. 
In short, the contributions of this work are summarized as follows:
\begin{itemize}
    \item We propose a lightweight static malware detection pipeline that combines ordinal-aware Windows API import extraction with efficient sequence-based deep learning.

     \item We introduce a decoupled post-hoc behavioral interpretation layer grounded in the MITRE ATT\&CK framework, which enhances analyst-oriented explainability without influencing classification decisions or degrading inference efficiency.
     
     \item We developed Delphi Scanner as a fully functional desktop application, and evaluate its detection performance on a real-world malware dataset, demonstrating competitive accuracy with a minimal computational footprint.
\end{itemize}

The remainder of this paper is organized as follows. Section~\ref{sec:related_work} reviews related work, Section~\ref{sec:System} describes the proposed system, Section~\ref{sec:Experiments} presents the experimental setup, Section~\ref{sec:Results} reports the results. Section~\ref{sec:case} shows a case study of the proposed system, Section~\ref{sec:Discussion and Limitations} discusses limitations and implications, and Section~\ref{sec:conclusion} concludes the paper.

\section{Related Work}
\label{sec:related_work}
Research on malware detection has progressed from signature-centric pipelines toward learning-based systems that exploit both static and dynamic evidence. This section reviews recent advances most closely related to our work.

\subsection{API-centric representations: dynamic call sequences and static imports}
Dynamic API call sequence modeling has been extensively explored using recurrent and convolutional architectures, often treating sequences as a language-like signal 
. For instance, API-MalDetect uses an NLP-inspired encoder together with convolutional and recurrent components to learn from long API call traces \cite{maniriho2023apimaldetect}. Similarly, hybrid sequence models have been proposed to improve generalization and sample efficiency; Owoh \emph{et al.} combine GRUs with GAN-based augmentation to enhance detection performance on API call sequences while controlling computational overhead \cite{owoh2024grugan}. While dynamic traces require controlled execution environments and are sensitive to sandbox fidelity, many practical systems prefer static analogues of behavioral evidence. Static import-based analysis supports pre-execution screening and complements header/section features. Recent work has also explored API-oriented rule generation and signature assistance: APIARY proposes an API-driven automatic rule generator for YARA, aiming to produce discriminative signatures that can leverage API patterns originating from either static or dynamic sources \cite{coscia2025apiary, botcloud2015}. Such directions align with the broader trend of using API-level signals to bridge human analyst intuition and automated detection~\cite{gaussian_mixture, 2023cyberattacks}.

\subsection{Deep learning directly over binaries}
Beyond engineered feature vectors, end-to-end models have attempted to learn directly from raw executables. MalConv demonstrated the feasibility of byte-level convolutional modeling over large PE inputs, catalyzing a substantial line of follow-on work \cite{raff2017malconv}. Subsequent improvements addressed scalability and representation capacity. Notably, Raff \emph{et al.} proposed constant-memory sequence classification to remove input-length constraints and improve training efficiency (often referred to as MalConv2 in the literature and associated tooling) \cite{raff2020constantmemory}. These models reduce reliance on hand-crafted features but raise challenges in interpretability, susceptibility to spurious correlations, and robustness to distribution shift or deliberate manipulation of input bytes.

Transformer-based approaches have also expanded rapidly in malware and binary analysis. I-MAD introduced a transformer architecture (Galaxy Transformer) for representing executable semantics at multiple granularities and paired it with an interpretable classification component \cite{li2021imad}. In addition, surveys highlight the breadth of transformer usage across malicious software detection settings, including byte/opcode modeling and hybrid representations, while emphasizing open issues around efficiency, dataset bias, and reproducibility \cite{alshomrani2024transformersurvey}. A recent SoK further systematizes the landscape of transformer-based malware analysis, highlighting design choices (tokenization, context length, pretraining, and multimodality) and stressing evaluation pitfalls under evolving threat conditions \cite{perry2024soktransformers}.

\subsection{Graph-based and relational static representations}
A complementary direction models executables as structured objects (graphs) rather than sequences or flat feature vectors. Graph learning can capture relations between heterogeneous static features (e.g., imports co-occurring with section/string attributes) and has been explored to improve robustness under drift. MFGraph, for instance, constructs feature graphs from static PE attributes and applies graph convolution to learn representations that remain comparatively stable under temporal changes \cite{zou2024mfgraph}. Such approaches suggest that explicitly modeling feature relations may provide resilience when distributions shift due to evolving development toolchains, packing practices, or malware family turnover.

\subsection{Adversarial machine learning}
Adversarial machine learning has likewise emerged as a practical threat model for both feature-based and raw-byte detectors. GAMBD proposes a gradient-based approach to generate functionality-preserving adversarial malware against MalConv, illustrating that even strong deep detectors can be evaded via constrained PE modifications \cite{li2023gambd}. Defensive research has responded with mechanisms that explicitly manage or scan the functionality-preserving attack space. Liu \emph{et al.} propose attack-space management to harden byte-sequence malware detection, aiming to reduce the need for repeated retraining and to improve resilience against both white-box and black-box attacks \cite{liu2024attackspacemgmt}. At a higher level, surveys of PE malware evasion methods emphasize that transformation-, concealment-, and attack-based strategies are routinely combined in practice, challenging detectors that rely on narrow assumptions about observable structure \cite{geng2024evasionSurvey}. These findings motivate approaches that (i) prioritize robust static signals, (ii) evaluate under realistic evasion settings, and (iii) maintain operational feasibility.

In summary, prior work spans engineered static PE feature modeling \cite{anderson2018ember,bakerdelaguila2024lowparam}, dynamic API-call sequence learning \cite{maniriho2023apimaldetect,owoh2024grugan}, end-to-end byte and transformer-based detectors \cite{raff2017malconv,raff2020constantmemory,li2021imad,alshomrani2024transformersurvey,perry2024soktransformers}, and robustness studies under packing and adversarial manipulation \cite{li2023gambd,liu2024attackspacemgmt,geng2024evasionSurvey}. Our work builds on these insights while focusing on an API-centric static representation tailored for efficient PE analysis and evaluated under realistic constraints, thereby complementing sequence-heavy dynamic approaches and resource-intensive end-to-end models.

\section{Delphi Scanner System}
\label{sec:System}  
This section presents Delphi Scanner, a static malware detection system for local, interpretable, and low-latency analysis of Windows PE files. The system integrates machine learning based classification with rule-based behavioral interpretation to deliver both an automated malware verdict and analyst-oriented explanatory insights. By operating exclusively on static binary features and avoiding code execution, Delphi Scanner enables safe and efficient malware screening suitable for end-user and analyst environments.

\begin{figure*}
    \centering
    \includegraphics[width=0.8\linewidth]{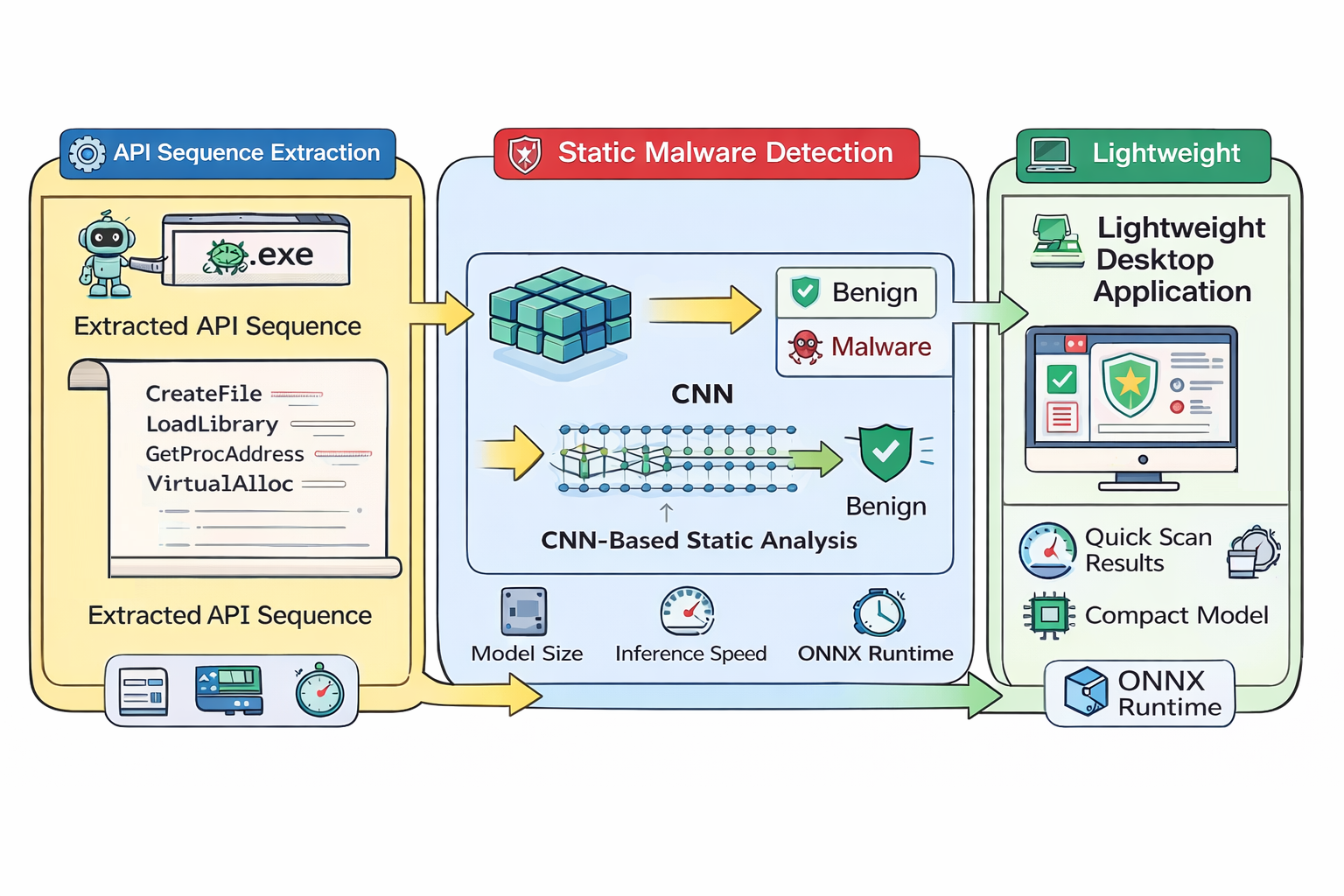}
    \caption{High-level architecture of Delphi Scanner: static feature extraction from the PE Import Address Table, CNN-based malware classification, rule-based behavioral interpretation layer operating in parallel with the CNN, and result presentation in the desktop interface.}

    \label{fig:system_architecture}
\end{figure*}

\subsection{System Overview}
Delphi Scanner targets four operational requirements: local and safe analysis without executing untrusted binaries, low-latency inference for interactive malware triage, interpretable behavioral output, and lightweight deployment with minimal runtime overhead.

Figure~\ref{fig:system_architecture} presents the high-level architecture. Given an input PE file, the system processes the binary through a modular pipeline comprising four components: static feature extraction, machine learning inference, behavioral interpretation, and user-facing result presentation. Each component operates independently and exchanges data through well-defined representations, facilitating modularity and extensibility.

\begin{itemize}
    \item Static Feature Extraction: The input PE file is statically parsed to extract imported Windows API functions from the Import Address Table (IAT). Ordinal-based imports are resolved where possible, and auxiliary static indicators, including section entropy and dynamic loading patterns, are computed. The resulting API sequence is encoded into a fixed-length numerical representation.
    
    \item Machine Learning Inference: The encoded feature representation is processed by a CNN that estimates the probability of maliciousness. Inference is optimized for low-latency execution and is performed independently of any handcrafted behavioral rules.

    \item Behavioral Interpretation: In parallel with classification, extracted APIs are mapped to predefined malicious capability categories, such as network communication, registry persistence, and process manipulation. This rule-based layer produces human-readable behavioral indicators without influencing the classifier’s prediction.
    
    \item User Interface: The outputs of the classification and interpretation stages are aggregated and presented through a desktop graphical interface, including the final verdict, confidence score, and detected behavioral indicators.
    
\end{itemize}

This architecture enforces a strict separation between statistical detection and behavioral interpretation, enabling accurate predictions while providing transparent explanatory context.
 
\subsection{Static Feature Extraction Module}
\label{sec:Static}  
This module describes the static analysis pipeline used by Delphi Scanner to transform a Windows PE file into a numerical representation suitable for machine learning--based classification. The extraction process targets semantically meaningful behavioral signals while remaining lightweight and resilient to common obfuscation techniques.

\subsubsection{PE Parsing and Import Extraction}

Delphi Scanner parses the binary structure of Windows PE files to extract imported Windows API functions from the Import Address Table (IAT). The extraction follows a seven-step pipeline, described below.

\begin{itemize}

\item Load and validate: The binary is read and validated by checking the DOS header magic bytes (\texttt{MZ}) and the PE signature (\texttt{PE\textbackslash0\textbackslash0}). Files exceeding 100 MB are rejected prior to parsing to bound memory usage.      

\item Locate the Import Directory: The parser navigates the PE Optional Header to the Data Directories array and retrieves the Import Table entry, which points to the Import Directory Table in the binary's virtual address space.

\item Iterate DLLs in import order: Each \texttt{IMAGE\_\allowbreak IMPORT\_\allowbreak DESCRIPTOR} structure yields the name of a dependent DLL (e.g., \texttt{KERNEL32.dll}).

\item Extract function names: For each DLL, the associated Import Lookup Table (ILT) is traversed entry by entry. Each entry is either a name reference, containing a function name string (e.g., \texttt{CreateFileW}), or an ordinal entry with a 16-bit ordinal number.

\item Resolve ordinal imports: Ordinal-based imports are resolved using the OrdinalResolver module, which contains built-in lookup tables for eight major Windows system libraries: \texttt{kernel32.dll}, ntdll.dll, \texttt{ws2\_\allowbreak 32.dll}, \texttt{user32.dll}, \texttt{advapi32.dll}, \_\allowbreak \texttt{shell32.dll}, \texttt{wininet.dll}, and \texttt{urlmon.dll}, covering approximately 300 function mappings. For example, \texttt{kernel32.dll} ordinal 581 resolves to GetProcAddress. Ordinals that cannot be resolved against the built-in tables are mapped to the special \texttt{<UNK>} token.

\item Construct the ordered sequence: Resolved function names are appended to the sequence in DLL order, then in function order within each DLL.

\item Vectorize: Each API name is mapped to an integer index via a fixed vocabulary of 10,000 tokens. Sequences are truncated to a maximum length of 512 tokens; shorter sequences are padded with a dedicated \texttt{<PAD>} token to produce uniform-length inputs for the convolutional classifier.

\end{itemize}

\subsubsection{API Sequence Construction}

Following import extraction and ordinal resolution, the system constructs an ordered sequence of API identifiers representing the static capabilities of the executable. Each API name is mapped to a unique integer index using a predefined vocabulary generated during training. To enable batch processing and fixed-size model inputs, sequences are normalized to a uniform length by padding shorter sequences and truncating longer ones. This normalization ensures compatibility with convolutional neural network architectures while preserving the most informative portion of the API sequence.

\begin{algorithm}[t]
\caption{Delphi Scanner Detection Pipeline}
\label{alg:delphi_scanner_pipeline}
\begin{algorithmic}[1]
\Require PE file $f$, API vocabulary $\mathcal{V}$, max sequence length $L$, decision threshold $\tau$, rule set $\mathcal{R}$
\Ensure Verdict $v$, confidence $p$, behavioral indicators $\mathcal{B}$

\State $pe \gets \textsc{ParsePE}(f)$
\State $(A, M) \gets \textsc{ExtractImports}(pe)$
\Comment{$A$: ordered API list, $M$: metadata (entropy, ordinals, etc.)}

\State $A \gets \textsc{ResolveOrdinals}(A)$
\State $s \gets \textsc{BuildSequence}(A, \mathcal{V})$
\State $x \gets \textsc{NormalizeSequence}(s, L)$
\Comment{pad/truncate to fixed length $L$}

\State $p \gets \textsc{CNNInfer}(x)$
\Comment{maliciousness probability}

\State $\mathcal{B} \gets \textsc{InterpretBehaviors}(A, \mathcal{R})$
\Comment{map APIs to capability categories + severity}

\If{$p \ge \tau$}
    \State $v \gets \text{"Malware"}$
\Else
    \State $v \gets \text{"Benign"}$
\EndIf

\State \Return $(v, p, \mathcal{B})$
\end{algorithmic}
\end{algorithm}

\subsubsection{Auxiliary Static Indicators}

In addition to API sequences, Delphi Scanner computes auxiliary static indicators that provide contextual information about the reliability and limitations of static analysis. These indicators are not used as direct inputs to the classification model but are exposed to the user to support result interpretation:

\begin{itemize}
    \item \textbf{Entropy:} The Shannon entropy of the executable, used as a coarse indicator of packing or compression that may limit the completeness of static import analysis.
    \item \textbf{Dynamic Loading Indicators:} The presence of APIs associated with runtime function resolution, which may enable evasion of import-based detection.
    \item \textbf{API Count:} The total number of extracted APIs, reflecting the richness of the available static feature set.
\end{itemize}

The output of the static feature extraction stage consists of a fixed-length numerical representation of the API sequence for machine learning inference, together with auxiliary metadata describing the extraction process, including entropy values, ordinal resolution statistics, and dynamic loading indicators. Feature extraction is fully decoupled from classification, preserving modularity and facilitating future extensions.

\subsection{Malware Classification Model}
\label{sec:Classification}
This subsection describes the malware classification model used in Delphi Scanner to estimate the maliciousness of a given PE file, corresponding to the \textsc{CNNInfer} step (line~6) of Algorithm~\ref{alg:delphi_scanner_pipeline}. The task is formulated as a binary classification problem in which each input sample is represented as an ordered sequence of Windows API identifiers, as described in Section~\ref{sec:Static}. The model produces a probabilistic estimate of maliciousness $p \in [0, 1]$, where $p$ denotes the softmax output probability that the input belongs to the malware class (i.e., $p = \text{softmax}(\text{logits})[1]$). This score is compared against a decision threshold $\tau$ (default $\tau = 0.5$): samples with $p \geq \tau$ are classified as malware, and those with $p < \tau$ as benign, as detailed in Algorithm~\ref{alg:delphi_scanner_pipeline}.

\subsubsection{Model Architecture}
\begin{figure*}[t]
    \centering
    \includegraphics[width=0.5\linewidth]{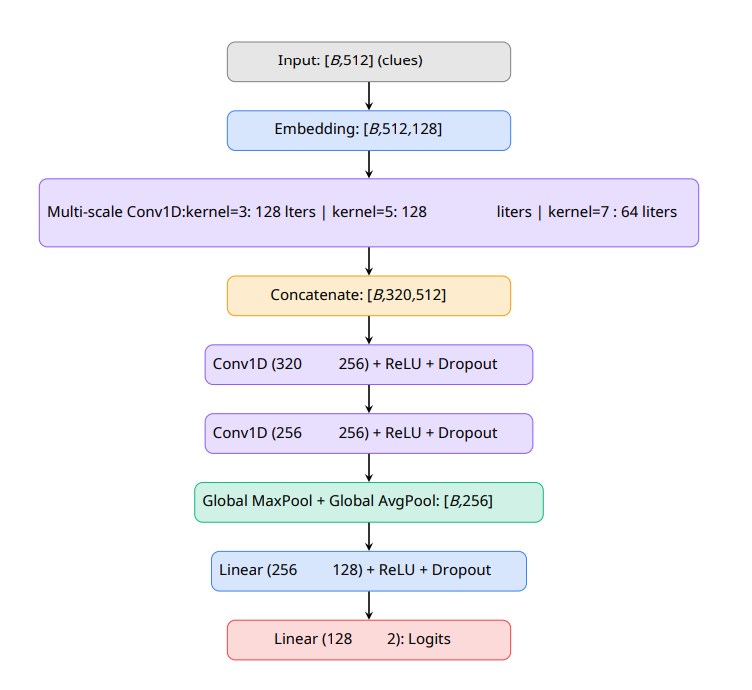}
    \caption{Architecture of the 1D CNN classifier, combining multi-scale convolutional filters (kernels 3, 5, 7) with global pooling to produce a fixed-length representation for binary malware classification.}
    \label{fig:CNN_model}
\end{figure*}

Delphi Scanner employs a convolutional neural network (CNN) selected for its favorable balance between detection performance and computational efficiency in static API sequence modeling. The deployed architecture, shown in Figure~\ref{fig:CNN_model}, processes API sequences of fixed length $512$. Each API identifier is mapped to a $128$-dimensional embedding, yielding an embedded sequence of shape $(512, 128)$. The embedded representation is passed through a multi-scale one-dimensional convolutional module comprising parallel convolutional kernels of sizes $3$, $5$, and $7$, with $128$, $128$, and $64$ filters, respectively, enabling the extraction of local patterns at multiple temporal scales. The resulting feature maps are concatenated and further processed by two additional Conv1D layers with ReLU activation and dropout for regularization. Global max pooling and global average pooling are applied to obtain a fixed $256$-dimensional representation, which is subsequently fed into a fully connected layer followed by a final linear layer to produce the logits for binary classification.

\subsubsection{Model Deployment}
For efficient and portable inference, the trained CNN model is exported to the Open Neural Network Exchange (ONNX) format and executed using the ONNX Runtime. This deployment approach decouples model training from inference and enables direct execution within the Rust-based backend without reliance on heavyweight machine learning frameworks. As a result, Delphi Scanner maintains a compact runtime environment while achieving sub-millisecond inference latency on standard hardware, supporting responsive malware analysis in desktop deployment scenarios.

\subsection{Behavioral Interpretation Layer}
The behavioral interpretation layer provides explanatory context for malware detection results derived from the MITRE ATT\&CK framework~\cite{mitre_attack} and grounded in established malware analysis literature~\cite{arp2022and}. Each category is mapped to a corresponding technique identifier, as listed in the table above. Its role operates in parallel with the malware classification pipeline; it does not influence the final malware verdict and serves exclusively as an explanatory component. 

This layer maps observed API functions to predefined behavioral categories, each representing a class of malicious capability observed in real-world malware. Table~\ref{tab:api_behavior_mapping} summarizes the implemented categories, their trigger conditions, and the corresponding API patterns used to identify them. Each category is activated when a minimum number of representative APIs is detected. Per-category thresholds are calibrated to reduce false activations caused by incidental API usage common in legitimate software. A qualitative severity level (e.g., \textit{High} or \textit{Medium}) is then assigned based on the sensitivity of the matched capability.

\begin{table*}[t]
\centering
\caption{Behavioral categories, trigger conditions, severity levels, descriptions, associated API indicators, and MITRE ATT\&CK identifiers implemented in the interpretation layer.}
\label{tab:api_behavior_mapping}
\renewcommand{\arraystretch}{1.2}
\begin{tabular}{p{2.4cm} p{2.0cm} p{1.4cm} p{4.2cm} p{4.2cm} p{1.6cm}}
\hline
\textbf{Category} & \textbf{Trigger Condition} & \textbf{Severity} & \textbf{Description} & \textbf{API Indicators} & \textbf{ATT\&CK ID} \\
\hline
Process Injection
& $\geq$2 APIs matched & Critical
& APIs for code injection into other processes
& \texttt{VirtualAllocEx}, \texttt{WriteProcessMemory}, \texttt{CreateRemoteThread}, \texttt{NtCreateThreadEx}
& T1055 \\

Credential Access
& $\geq$1 API matched & Critical
& APIs for stealing credentials
& \texttt{CredEnumerateA/W}, \texttt{CryptUnprotectData}, \texttt{CryptDecrypt}
& T1555 \\

Keylogging
& $\geq$2 APIs matched & Critical
& APIs for keyboard monitoring
& \texttt{SetWindowsHookExA/W}, \texttt{GetAsyncKeyState}, \texttt{GetKeyboardState}
& T1056.001 \\

Registry Persistence
& $\geq$1 API matched & High
& APIs for establishing persistence via registry
& \texttt{RegSetValueExA/W}, \texttt{RegCreateKeyExA/W}
& T1547.001 \\

Network Communication
& $\geq$2 APIs matched & High
& APIs for network/C2 communication
& \texttt{WSAStartup}, \texttt{socket}, \texttt{connect}, \texttt{send}, \texttt{recv}, 
  \texttt{InternetOpenA/W}, \texttt{URLDownloadToFileA/W}
& T1071 \\

Service Manipulation
& $\geq$2 APIs matched & High
& APIs for Windows service manipulation
& \texttt{CreateServiceA/W}, \texttt{OpenServiceA/W}, \texttt{StartServiceA/W}, 
  \texttt{ControlService}
& T1543.003 \\

Anti-Debugging
& $\geq$1 API matched & Medium
& APIs to detect or evade debugging
& \texttt{IsDebuggerPresent}, \texttt{CheckRemoteDebuggerPresent}, 
  \texttt{NtQueryInformationProcess}
& T1622 \\
\hline
\end{tabular}
\end{table*}

\subsection{User Interface}
The user interface serves as the interaction layer between the analyst and the static analysis backend, aggregating the outputs of malware classification and behavioral interpretation into a unified view. For each analyzed PE file, the interface displays the malware verdict and associated confidence score, along with relevant static analysis metadata such as file characteristics, entropy values, and API counts. Behavioral threat indicators produced by the interpretation layer are grouped by category and visually differentiated by severity to support rapid malware triage. All analysis is performed locally, and the interface operates without reliance on external services, preserving user privacy and enabling low-latency interaction.

\begin{table}[t]
\caption{Technology stack used in the implementation of Delphi Scanner.}
\label{tab:tech_stack}
\centering
\resizebox{\columnwidth}{!}{%
\begin{tabular}{lll}
\hline
\textbf{Layer} & \textbf{Technology} & \textbf{Justification} \\
\hline
\multirow{3}{*}{Frontend}
 & React 19         & Reactive components, large ecosystem \\
 & TypeScript       & Static typing, improved maintainability \\
 & Tailwind CSS v4  & Utility-first styling, modern UI \\
\hline
\multirow{4}{*}{Backend}
 & Rust 1.70+       & Memory safety, native performance \\
 & Tauri v2         & Lightweight desktop framework \\
 & Goblin           & Pure Rust PE parsing, no C FFI \\
 & ONNX Runtime 1.22& Optimized inference, GPU support \\
\hline
\multirow{3}{*}{Training}
 & Python 3.10+     & Mature ML ecosystem \\
 & PyTorch 2.0      & Flexible training and debugging \\
 & scikit-learn     & Preprocessing and metrics \\
\hline
\end{tabular}}
\end{table}

\subsection{Implementation Overview}
Delphi Scanner is implemented as a lightweight desktop application following a modular client--backend architecture. The backend performs static analysis, feature extraction, machine learning inference, and behavioral interpretation, while the frontend provides an interface for initiating analysis and presenting results. The technology stack is summarized in Table~\ref{tab:tech_stack}. The backend is implemented in Rust to ensure performance and memory safety, and the frontend is developed using React and TypeScript with Tailwind CSS for consistent styling. Frontend and backend components are integrated using the Tauri framework, which enables secure communication between web-based interfaces and native system functionality.

\begin{figure*}[ht]
\centering
\resizebox{\textwidth}{!}{%
\begin{tikzpicture}[
    node distance = 0.6cm and 1.0cm,
    box/.style      = {rectangle, rounded corners=4pt, draw=black!70, fill=blue!8,
                       text width=2.4cm, align=center, minimum height=0.95cm, font=\small},
    trainbox/.style = {rectangle, rounded corners=4pt, draw=black!70, fill=orange!20,
                       text width=2.4cm, align=center, minimum height=0.95cm, font=\small},
    deploybox/.style= {rectangle, rounded corners=4pt, draw=black!70, fill=green!15,
                       text width=2.4cm, align=center, minimum height=0.95cm, font=\small},
    groupbox/.style = {rectangle, draw=gray!60, dashed, fill=gray!5,
                       inner sep=10pt, rounded corners=5pt},
    arrow/.style    = {-{Latex[length=2.5mm]}, thick, draw=black!80},
    darrow/.style   = {-{Latex[length=2.5mm]}, thick, dashed, draw=gray!60}
]

%% ── ROW 1: SHARED PREPROCESSING ──────────────────────────────────────────
\node[box] (pe)     {Raw PE\\File};
\node[box, right=of pe]    (parse)  {Validate \&\\Parse PE};
\node[box, right=of parse] (iat)    {Extract IAT \&\\Resolve Ordinals};
\node[box, right=of iat]   (seq)    {Ordered API\\Sequence};
\node[box, right=of seq]   (vec)    {Vectorize \&\\Pad ($L{=}512$)};

%% ── ROW 2: TRAINING BRANCH ────────────────────────────────────────────────
\node[trainbox, below=2.2cm of parse] (split) {Train / Test\\Split (80/20)};
\node[trainbox, right=of split]       (train) {Train CNN\\(PyTorch)};
\node[trainbox, right=of train]       (onnx)  {Export to\\ONNX};

%% ── ROW 3: INFERENCE BRANCH ───────────────────────────────────────────────
\node[deploybox, below=2.2cm of onnx]  (infer) {ONNX Runtime\\Inference};
\node[deploybox, left=of infer]         (rules) {Behavioral\\Rule Engine};
\node[deploybox, left=of rules]         (out)   {Verdict +\\Confidence +\\Indicators};

%% ── PREPROCESSING ARROWS ──────────────────────────────────────────────────
\draw[arrow] (pe)    -- (parse);
\draw[arrow] (parse) -- (iat);
\draw[arrow] (iat)   -- (seq);
\draw[arrow] (seq)   -- (vec);

%% ── vec → split: straight down then left ─────────────────────────────────
\draw[arrow] (vec.south) -- ++(0,-0.7) -| (split.north);

%% ── TRAINING ARROWS ───────────────────────────────────────────────────────
\draw[arrow] (split) -- (train);
\draw[arrow] (train) -- (onnx);
\draw[arrow] (onnx)  -- (infer);

%% ── INFERENCE ARROWS ──────────────────────────────────────────────────────
\draw[arrow] (infer) -- (rules);
\draw[arrow] (rules) -- (out);

%% ── DASHED (a): seq → rules  (API names for rule engine) ─────────────────
%% Route cleanly: seq.south → down left of training box → rules.north
\draw[darrow]
    (seq.south)
    -- ++(0, -0.35)
    -- ++(-0.4, 0)
    |- (rules.west);

%% ── DASHED (b): vec → infer  (deployment-time vectorized input) ──────────
%% Route: vec → right side of diagram → down → infer.east
\draw[darrow]
    (vec.east)
    -- ++(0.55, 0)
    |- (infer.east);

%% ── LEGEND (small, bottom-left) ───────────────────────────────────────────
\node[font=\scriptsize, anchor=west] at ([xshift=-0.2cm, yshift=-0.9cm]out.south west)
    {\tikz\draw[darrow,thick] (0,0) -- (0.6,0); \; dashed = secondary data flow};

%% ── BOUNDING GROUP BOXES ──────────────────────────────────────────────────
\begin{scope}[on background layer]
  \node[groupbox, fit=(pe)(parse)(iat)(seq)(vec),
        label={[font=\small\bfseries]above:Shared Preprocessing}] {};

  \node[groupbox, fit=(split)(train)(onnx),
        label={[font=\small\bfseries]above:Offline Training}] {};

  \node[groupbox, fit=(infer)(rules)(out),
        label={[font=\small\bfseries]above:Deployment-Time Inference}] {};
\end{scope}

\end{tikzpicture}
}

\caption{End-to-end pipeline of Delphi Scanner: shared preprocessing (blue), offline training (orange), and deployment-time inference (green). Dashed arrows indicate secondary data flows to the behavioral rule engine and classifier input.}
\label{fig:pipeline_flowchart}
\end{figure*}
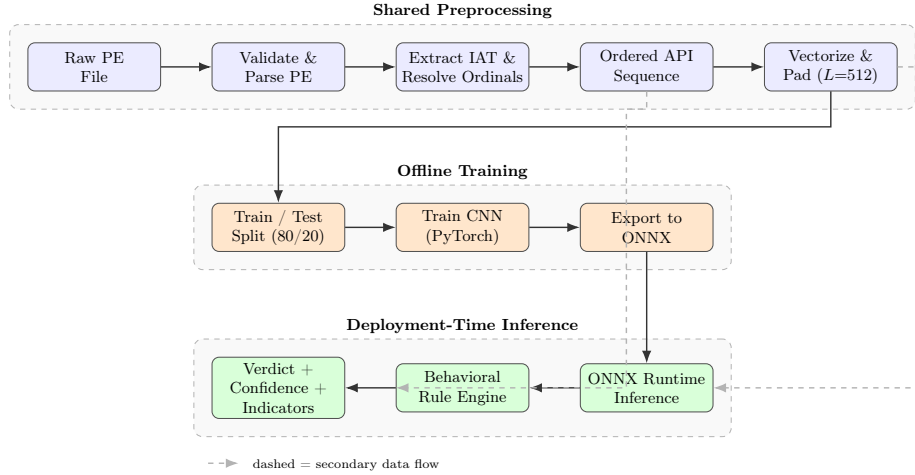

Communication between the frontend and backend is performed via a command-based interface, supporting structured analysis requests and dynamic rendering of results. All analysis is executed locally without dependence on external services, ensuring privacy preservation and low-latency operation. This modular design decouples analysis logic from presentation concerns, improving maintainability and facilitating future extensions, including the integration of additional static features, alternative classification models, or enhanced visualization components. The complete source code of the application, including backend, frontend, and model deployment logic, is publicly available as an open-source repository\footnote{\url{https://github.com/0xf1d0/delphi_scanner}}.Figure~\ref{fig:pipeline_flowchart} consolidates the complete Delphi Scanner pipeline across both offline training and deployment-time inference stages described in this section.

\section{Experiments}
\label{sec:Experiments}  

\subsection{Dataset Description and Data Preprocessing}
This study employs a large-scale dataset of Windows Portable Executable (PE) files constructed by combining a widely used public reference dataset with additional complementary sources. The goal of this aggregation is to increase sample diversity, reduce dataset bias, and improve the robustness of the trained models.

\subsubsection{Dataset Description}
The primary data source is the \textit{PE Malware Machine Learning Dataset} released by Practical Security Analytics~\cite{psa_dataset}, which contains real-world PE binaries labeled as malicious or benign. The dataset includes executables and dynamic-link libraries collected from multiple sources and time periods, with samples provided as complete binaries rather than pre-extracted features, enabling flexible static analysis.

To enrich the dataset and improve coverage of recent malware families and clean system binaries, additional data sources were integrated. Malware samples were obtained from publicly available repositories, including VirusShare~\cite{virusshare}, while benign samples were collected from legitimate Windows system binaries. In addition, a curated dataset containing pre-extracted API sequences was incorporated to increase variability in observed malware behaviors~\cite{maniriho2023apimaldetect}. The composition of all data sources and the resulting dataset statistics before and after deduplication are summarized in Table~\ref{tab:dataset_characteristics}. After aggregation, the combined dataset was deduplicated using SHA-256 cryptographic hashes to remove identical binaries and prevent data leakage between training and evaluation phases. This step reduced the dataset to 200,000 unique samples, as reported in Table~\ref{tab:dataset_characteristics}.

\begin{table}[t]
\centering
\caption{Dataset characteristics.}
\label{tab:dataset_characteristics}
\renewcommand{\arraystretch}{1.1}

% -------- Section 1 --------
\begin{tabular}{lrrr}
\multicolumn{4}{c}{\textbf{Datasets Sources}} \\
\hline
Source & Malware & Benign & Total \\
\hline
PE Malware ML Dataset & 114,737 & 86,812 & 201,549 \\
VirusShare & 20,000 & 0 & 20,000 \\
MalBehavD-V1 & 8,000 & 2,000 & 10,000 \\
Windows System32 & 0 & 5,000 & 5,000 \\
\hline
\end{tabular}

\vspace{1em}

% -------- Section 2 --------
\begin{tabular}{lrrr}
\multicolumn{4}{c}{\textbf{Combined Dataset Before and After Deduplication}} \\
\hline
Source & Malware & Benign & Total \\
\hline
Before deduplication & 142,737 & 93,812 & 236,549 \\
After deduplication & 120,000 & 80,000 & 200,000 \\
\hline
\end{tabular}

\end{table}

\subsubsection{Preprocessing Pipeline}
All PE files were processed using the static feature extraction pipeline described in Section~\ref{sec:Static}. Imported Windows API functions were extracted, ordinal-based imports were resolved when possible, and API sequences were transformed into fixed-length numerical representations suitable for machine learning inference. Samples with corrupted PE structures and insufficient API coverage were excluded. During preprocessing, an additional 9,636 samples were removed, resulting in a final dataset of 190,364 samples. The remaining samples were partitioned into training and test subsets using a fixed 80/20 split. Sequence normalization and dataset partitioning were performed prior to model training, with both subsets maintaining a balanced distribution of malicious and benign files.

\subsection{Algorithms}
To assess the effectiveness of different sequence modeling approaches for static malware detection, multiple deep learning architectures were implemented and evaluated in this study. All models operate on the same fixed-length API sequence representation described in Section~\ref{sec:Static} and are trained to perform binary classification, distinguishing malicious from benign PE files.

\subsubsection{Evaluated Architectures}
The evaluated models represent a range of sequence learning paradigms commonly employed in malware detection and sequential data analysis:

\begin{itemize}
    \item Convolutional Neural Networks (CNN): architectures designed to capture local patterns in API sequences using one-dimensional convolutional filters.
    \item Long Short-Term Memory (LSTM) networks: recurrent models capable of modeling long-range dependencies in sequential data.
    \item Gated Recurrent Units (GRU): a computationally efficient alternative to LSTMs with fewer parameters.
    \item Transformer-based models: attention-driven architectures that model global dependencies across the entire API sequence.
    \item Hybrid CNN--LSTM architectures: models that combine convolutional layers for local feature extraction with recurrent layers for sequential dependency modeling.
\end{itemize}

These architectures were selected to provide a comprehensive comparison of convolutional, recurrent, attention-based, and hybrid approaches.

\subsubsection{Training Configuration}
To ensure a fair and reproducible comparison, all models were trained using a unified training configuration. Hyperparameters were selected empirically based on convergence behavior, training stability, and validation performance, rather than exhaustive grid search, reflecting practical constraints commonly encountered in applied security research. The shared training configuration is summarized in Table~\ref{tab:training_config}.

All models were trained using cross-entropy loss with label smoothing to improve generalization and reduce prediction overconfidence. Training was terminated early when validation performance ceased to improve, ensuring efficient convergence while avoiding unnecessary overfitting. Although the models share identical input representations and training procedures, they differ in architectural complexity, parameter count, and computational characteristics, enabling a meaningful comparative analysis.

\begin{table}[t]
\centering
\caption{Common training configuration used for all evaluated models.}
\label{tab:training_config}
\begin{tabular}{ll}
\hline
\textbf{Hyperparameter} & \textbf{Value} \\
\hline
Batch size & 64 \\
Maximum epochs & 50 \\
Initial learning rate & $1 \times 10^{-3}$ \\
Weight decay & $1 \times 10^{-4}$ \\
Optimizer & AdamW \\
Scheduler & ReduceLROnPlateau \\
Early stopping patience & 5 epochs \\
\hline
\end{tabular}
\end{table}

\subsection{Evaluation Metrics}
All models were trained and evaluated under identical experimental conditions to ensure a fair and reproducible comparison. Specifically, the same dataset splits, preprocessing pipeline, and API sequence representations were used across all evaluated architectures. Model performance was assessed exclusively on a held-out test set that was not used during training. Detection effectiveness was evaluated using standard binary classification metrics commonly adopted in malware detection research:

\begin{itemize}
    \item Accuracy: defined as the proportion of correctly classified samples.
    \item Precision:  measuring the fraction of samples predicted as malicious that are truly malicious.
    \item Recall: measuring the fraction of malicious samples correctly identified.
    \item F1-score: computed as the harmonic mean of precision and recall.
\end{itemize}

In addition to classification performance, system-level metrics relevant to practical deployment were also measured. Inference latency was calculated as the average time required to process a single PE file during model execution, while model size was recorded to assess memory footprint. This evaluation protocol ensures that selected models are not only effective in detecting malware, but also feasible for real-world usage within the constraints of a local analysis environment.

\subsection{Implementation}
All experiments were implemented and executed in a controlled local environment to ensure reproducibility and consistent performance measurements. Model training and evaluation were conducted on a single workstation using a modern multi-core CPU and a dedicated GPU to accelerate deep learning workloads. The experimental setup is summarized in Table~\ref{tab:hardware_config}. The system is equipped with an AMD Ryzen 7 5800H processor and an NVIDIA RTX 3060 Mobile GPU with dedicated VRAM, providing sufficient computational resources for training deep learning models on large-scale PE datasets. The operating system used was Windows 11 Pro, which aligns with the target deployment environment of the proposed malware analysis tool.

Model training was implemented using the PyTorch deep learning framework, which was selected for its flexibility, extensive ecosystem, and ease of experimentation. CUDA support was enabled to leverage GPU acceleration during training, significantly reducing training time for recurrent and attention-based models. All preprocessing steps, including API sequence extraction and vectorization, were performed prior to training to minimize runtime overhead during model optimization. Inference experiments were conducted using the trained models exported to the ONNX format, enabling efficient execution through the ONNX Runtime. This separation between training and inference environments reflects the intended system design, where models are trained offline and deployed within a lightweight desktop application for real-time analysis.

\subsection{Robustness Evaluation}
To assess the robustness of the trained CNN model beyond the held-out test set, three complementary evaluation experiments were conducted, targeting generalization to unseen malware, resilience to executable packing, and resistance to adversarial manipulation of PE binary structures. All experiments were performed using the deployed ONNX model without retraining or threshold adjustment.

\subsubsection{Generalization to Unseen Malware}
To evaluate the model's ability to generalize beyond the training distribution, a set of 5,647 malware samples was collected from MalwareBazaar~\cite{malwarebazaar2024}, a publicly accessible malware repository aggregating recent and actively distributed threats. All samples were processed through the same static feature extraction pipeline. For this experiment, recall serves as the primary evaluation metric since all collected samples are labeled malicious.

\subsubsection{UPX Packing Robustness}
To evaluate the impact of executable packing on detection performance, 106 UPX-packed malware samples were collected and subjected to automated unpacking using the UPX decompressor, yielding a paired set of 106 packed and 106 unpacked binaries derived from identical malware instances. Both variants of each sample were independently analyzed through the static feature extraction pipeline. 

\subsubsection{Adversarial Manipulation Robustness}
To assess resistance to functionality-preserving adversarial perturbations of PE files, 300 adversarial samples were generated from malicious samples across three attack techniques, with 100 samples per technique. The evaluated techniques are as follows. 

\textit{API Benign Injection} appends legitimate, non-malicious API imports to the malware's import table to dilute the learned malicious API patterns without altering program functionality. 

\textit{Full DOS Padding} (FullDOS) extends the DOS header stub region of the PE file with arbitrary byte content, inflating the binary without modifying executable code or imports. 

\textit{Section Padding} appends benign byte sequences to existing PE sections, increasing file size while preserving the executable's original import structure and behavior.

\section{Results and Analysis}
\label{sec:Results}  

This section presents the experimental results obtained using the evaluation protocol described in the previous section. The objective is to assess the detection performance, class-wise behavior, and deployment feasibility of the evaluated models under identical experimental conditions.

\begin{table}[t]
\centering
\caption{Hardware and software configuration used for experiments.}
\label{tab:hardware_config}
\begin{tabular}{ll}
\hline
\textbf{Component} & \textbf{Specification} \\
\hline
Processor & AMD Ryzen 7 5800H (8 cores) \\
GPU & NVIDIA RTX 3060 Mobile (6 GB VRAM) \\
RAM & 16 GB DDR4 \\
Operating System & Windows 11 Pro \\
CUDA Version & 12.6 \\
PyTorch Version & 2.0+ \\
\hline
\end{tabular}
\end{table}

\begin{table*}[t]
\centering
\caption{Comparative benchmark of the evaluated architectures on the test set.}
\label{tab:model_comparison}
\begin{tabular}{lccccc}
\hline
\textbf{Model} & \textbf{Accuracy (\%)} & \textbf{F1-score} & \textbf{Inference Time (ms)} & \textbf{Model Size (MB)} \\
\hline
CNN        & 95.35 & 0.9536 & 0.56 & 1.53 \\
GRU        & 95.23 & 0.9524 & 0.30 & 1.79 \\
Transformer & 95.04 & 0.9505 & 1.39 & 1.60 \\
LSTM       & 94.95 & 0.9496 & 0.33 & 1.98 \\
CNN--LSTM  & 94.82 & 0.9483 & 0.89 & 1.72 \\
\hline
\end{tabular}
\end{table*}

\subsection{Overall Detection Performance}
The overall detection performance of the evaluated architectures is summarized in Table~\ref{tab:model_comparison}. All models achieve high accuracy and F1-score on the held-out test set, confirming that static API sequence–based analysis is effective for malware detection. Among the evaluated approaches, the CNN-based model achieves the best trade-off between detection performance and efficiency. It achieves the highest accuracy (95.35\%) and F1 Score (0.9536), while maintaining a compact model size and low inference latency. Recurrent and transformer-based models exhibit comparable accuracy but incur higher computational overhead, resulting in increased inference time and larger memory footprints.

\begin{figure*}[t]
\centering
\includegraphics[width=0.95\linewidth]{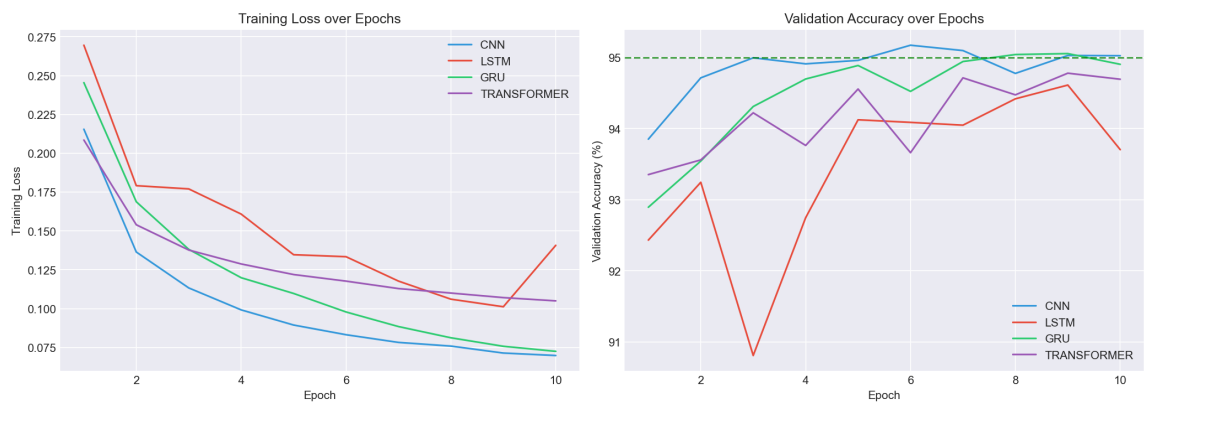}
\caption{Validation accuracy and loss over training epochs for all evaluated architectures.}
\label{fig:training_history}
\end{figure*}

Figure~\ref{fig:training_history} illustrates the complete training history of the evaluated models in terms of accuracy and loss. The CNN model demonstrates stable and rapid convergence, reaching optimal performance in fewer epochs compared to recurrent and transformer-based architectures. This behavior further supports the suitability of convolutional models for modeling local API patterns in static malware detection. Recurrent and transformer-based models exhibit slower convergence and higher variance during training, indicating increased optimization complexity compared to convolutional architectures.

\subsection{Per-Class Performance}
To further analyze the behavior of the selected CNN model, class-wise performance metrics were examined. Table~\ref{tab:cnn_class_metrics} reports precision, recall, and F1-score for both benign and malicious classes. The CNN model achieves balanced performance across both classes, with high precision and recall values for malware and benign samples alike. This balance indicates that the model does not exhibit a strong bias toward either class, limiting both false alarms and undetected malware, which is particularly important in malware detection scenarios where both false positives and false negatives can have practical consequences.

\subsection{Inference Efficiency and Model Size}
In addition to detection accuracy, inference latency and model size were evaluated to assess the practicality of deploying the models in a desktop malware analysis tool. As shown in Table~\ref{tab:model_comparison}, the CNN model achieves sub-millisecond inference time while maintaining a small memory footprint. Compared to recurrent and transformer-based architectures, the CNN model offers faster inference and reduced computational overhead, making it well-suited for interactive malware analysis scenarios. These characteristics are particularly important for local, real-time analysis environments where responsiveness and resource efficiency are critical.

\begin{table}[t]
\centering
\caption{Per-class performance metrics of the CNN model on the test set.}
\label{tab:cnn_class_metrics}
\begin{tabular}{lcccc}
\hline
\textbf{Class} & \textbf{Precision} & \textbf{Recall} & \textbf{F1-score} & \textbf{Support} \\
\hline
Benign (0)  & 0.9478 & 0.9602 & 0.9540 & 19,037 \\
Malware (1) & 0.9595 & 0.9468 & 0.9531 & 19,036 \\
\hline
Macro Avg   & 0.9537 & 0.9535 & 0.9536 & 38,073 \\
Weighted Avg & 0.9537 & 0.9535 & 0.9536 & 38,073 \\
\hline
\end{tabular}
\end{table}

\subsection{Error Analysis}
To further understand the behavior of the evaluated models beyond aggregate performance metrics, an error analysis was conducted focusing on false positive (FP) and false negative (FN) outcomes. In the context of malware detection, false negatives—malicious samples incorrectly classified as benign—are generally considered more critical than false positives, as they represent undetected threats that may execute without warning. Table~\ref{tab:error_analysis} summarizes the false positive and false negative counts and rates for all evaluated architectures. Across all models, false negative rates are consistently higher than false positive rates, reflecting the inherent difficulty of identifying stealthy or weakly represented malicious behaviors using static analysis alone. Among the evaluated approaches, the CNN-based model achieves the lowest false positive rate (2.0\%) and a comparatively low false negative rate (2.7\%), indicating a favorable balance between detection sensitivity and misclassification risk.

Recurrent and attention-based models, including LSTM, GRU, and Transformer architectures, exhibit slightly higher false negative rates, suggesting a reduced ability to generalize across diverse malware behaviors in this setting. Hybrid CNN--LSTM models show the highest error rates among the evaluated architectures, likely due to increased model complexity without corresponding gains in discriminative capability. Overall, the error analysis reinforces the suitability of the CNN-based architecture for deployment in security-sensitive environments, where minimizing undetected malware while maintaining acceptable false alarm rates is essential.

\begin{table}[t]
\caption{Error analysis by model showing false positive (FP) and false negative (FN) counts and rates.}
\label{tab:error_analysis}
\centering
\footnotesize
\setlength{\tabcolsep}{4pt}
\begin{tabular}{lcccc}
\hline
\textbf{Model} & \textbf{FP} & \textbf{FN} & \textbf{FP (\%)} & \textbf{FN (\%)} \\
\hline
CNN            & 758 & 1,013 & 2.0 & 2.7 \\
GRU            & 782 & 1,032 & 2.1 & 2.7 \\
Transformer    & 815 & 1,073 & 2.1 & 2.8 \\
LSTM           & 836 & 1,086 & 2.2 & 2.9 \\
CNN--LSTM      & 867 & 1,105 & 2.3 & 2.9 \\
\hline
\end{tabular}
\end{table}

\subsection{Performance Trade-off Analysis}
The results show a clear trade-off between model performance and computational efficiency. The CNN achieves the highest accuracy (95.35\%) and the lowest false positive rate (2.0\%), while maintaining a compact footprint (1.53 MB) and sub-millisecond inference latency, making it suitable for lightweight malware detection. In contrast, LSTM and GRU models require 1.5--3$\times$ more memory and yield higher false negative rates despite their ability to capture sequential dependencies. The Transformer model exhibits the highest inference cost (1.39 ms), approximately 2.5$\times$ slower than the CNN, without performance gains, indicating that local API co-occurrence patterns are more informative than global dependencies in static analysis.

The behavioral interpretation layer introduces negligible overhead due to its rule-based set-membership lookups and operates independently of the neural pipeline. As a result, explainability is effectively cost-free at inference time compared to post-hoc methods such as SHAP or LIME. The main runtime bottleneck is PE parsing and I/O, which takes approximately 50 ms per sample—several orders of magnitude higher than inference time—and can be mitigated through parallel processing. Training time scales approximately linearly with dataset size, with the full training process completing in about 11 minutes on an RTX 3060, supporting periodic retraining as new malware families emerge.

\subsection{Robustness Evaluation Results}
The results of the three robustness evaluation experiments are summarized in Table~\ref{tab:robustness}. Overall, the CNN model demonstrates consistent resilience across all evaluated conditions, maintaining high recall.

\begin{table*}[ht]
\centering
\caption{Robustness evaluation results across generalization, packing, and adversarial manipulation experiments.}
\label{tab:robustness}
\begin{tabular}{llc}
\hline
\textbf{Experiment} & \textbf{Condition} & \textbf{Recall (\%)} \\
\hline
Generalization      & MalwareBazaar (5,647 samples)   & 89.13 \\
\hline
\multirow{2}{*}{UPX Packing} 
                    & Packed (106 samples)            & 92.08 \\
                    & Unpacked (106 samples)          & 97.16 \\
\hline
\multirow{3}{*}{Adversarial Manipulation} 
                    & API Benign Injection            & 95.00 \\
                    & Full DOS Padding                & 97.85 \\
                    & Section Padding                 & 97.01 \\
\hline
\end{tabular}
\end{table*}

\subsubsection{Generalization to Unseen Malware.}
The model achieves a recall of 89.13\% on the MalwareBazaar sample set, demonstrating meaningful generalization to malware samples outside the training distribution. The observed reduction relative to the test-set malware recall of 94.68\% is consistent with the expected effect of distribution shift, as MalwareBazaar aggregates recently submitted and actively distributed threats that may employ API usage patterns underrepresented in the training data. Nevertheless, the result confirms that the model captures generalizable malicious API patterns rather than overfitting to dataset-specific artifacts.

\subsubsection{UPX Packing Robustness.}
The paired packing experiment reveals a measurable but moderate impact of UPX packing on detection recall. The model achieves 97.16\% recall on unpacked binaries, consistent with performance on the main test set, while recall on the corresponding packed variants decreases to 92.08\%, a reduction of 5.08 percentage points. This decline is attributable to the reduced import table visibility in packed executables, where the IAT exposes only the unpacking stub rather than the full malicious API sequence. Despite this degradation, the model retains substantial detection capability on packed samples, suggesting that packing-stub API patterns carry residual discriminative signal.

\subsubsection{Adversarial Manipulation Robustness.}
The model demonstrates strong resistance to all three evaluated adversarial manipulation techniques. Recall remains high under API Benign Injection (95.00\%), Full DOS Padding (97.85\%), and Section Padding (97.01\%), with no technique reducing recall below 95\%. These results indicate that the CNN classifier does not rely on superficial binary-level features susceptible to padding-based perturbations, and that the injection of benign API imports produces only a limited dilution effect on the learned malicious sequence representations. The relative robustness to API Benign Injection is particularly noteworthy, as this technique directly targets the feature space exploited by the model, yet results in only a modest recall reduction of 0.35 percentage points compared to the original test set performance.

\subsection{Summary of Results}
Overall, the experimental results demonstrate that convolutional architectures provide an effective and efficient solution for static malware detection based on API sequences. The CNN model achieves strong detection performance, balanced class-wise behavior, and low inference latency, validating its selection as the deployed classifier in the proposed system.

\section{Case Study: Delphi Scanner Output}
\label{sec:case}

To complement the quantitative evaluation presented in the previous section, we provide a qualitative case study illustrating how Delphi Scanner operates in practice and the type of outputs it produces for analysts. The objective of this case study is to demonstrate the system’s end-to-end behavior, including static feature extraction, machine learning–based classification, and rule-based behavioral interpretation, as exposed through the user interface. Three representative screenshots are used to illustrate (i) the main interface, (ii) the analysis of a benign executable, and (iii) the analysis of a malicious sample.
\begin{figure}[t]
    \centering
    \includegraphics[width=0.9\linewidth]{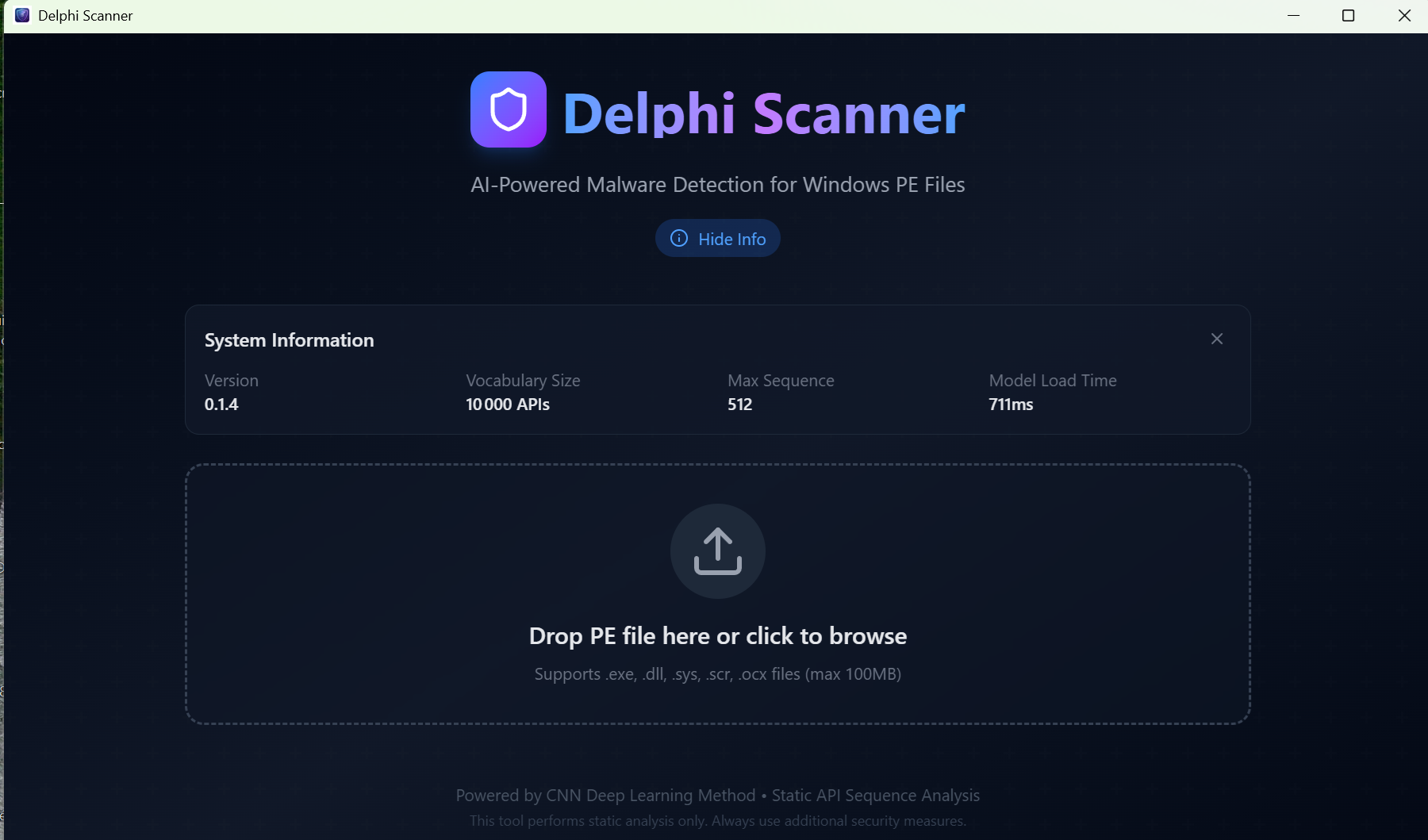}
    \caption{Delphi Scanner main user interface.}
    \label{fig:delphi-ui}
\end{figure}

\subsection{User Interface and System Overview}
Figure~\ref{fig:delphi-ui} presents the main user interface of Delphi Scanner. The application is designed as a lightweight desktop tool supporting fully offline analysis of Windows PE files. Upon launch, the interface displays system-level information, including the model version, API vocabulary size, maximum sequence length, and model load time. This metadata provides transparency regarding the underlying detection engine and reinforces the system’s emphasis on efficiency and local deployment.

The interface supports a simple drag-and-drop workflow for PE files, enabling rapid analysis without requiring external services or sandbox execution. All results are computed locally and displayed within a single consolidated view, combining the malware verdict, confidence score, static file metadata, and behavioral indicators.

\subsection{Analysis of a Benign File}
\begin{figure}[t]
    \centering
    \includegraphics[width=0.9\linewidth]{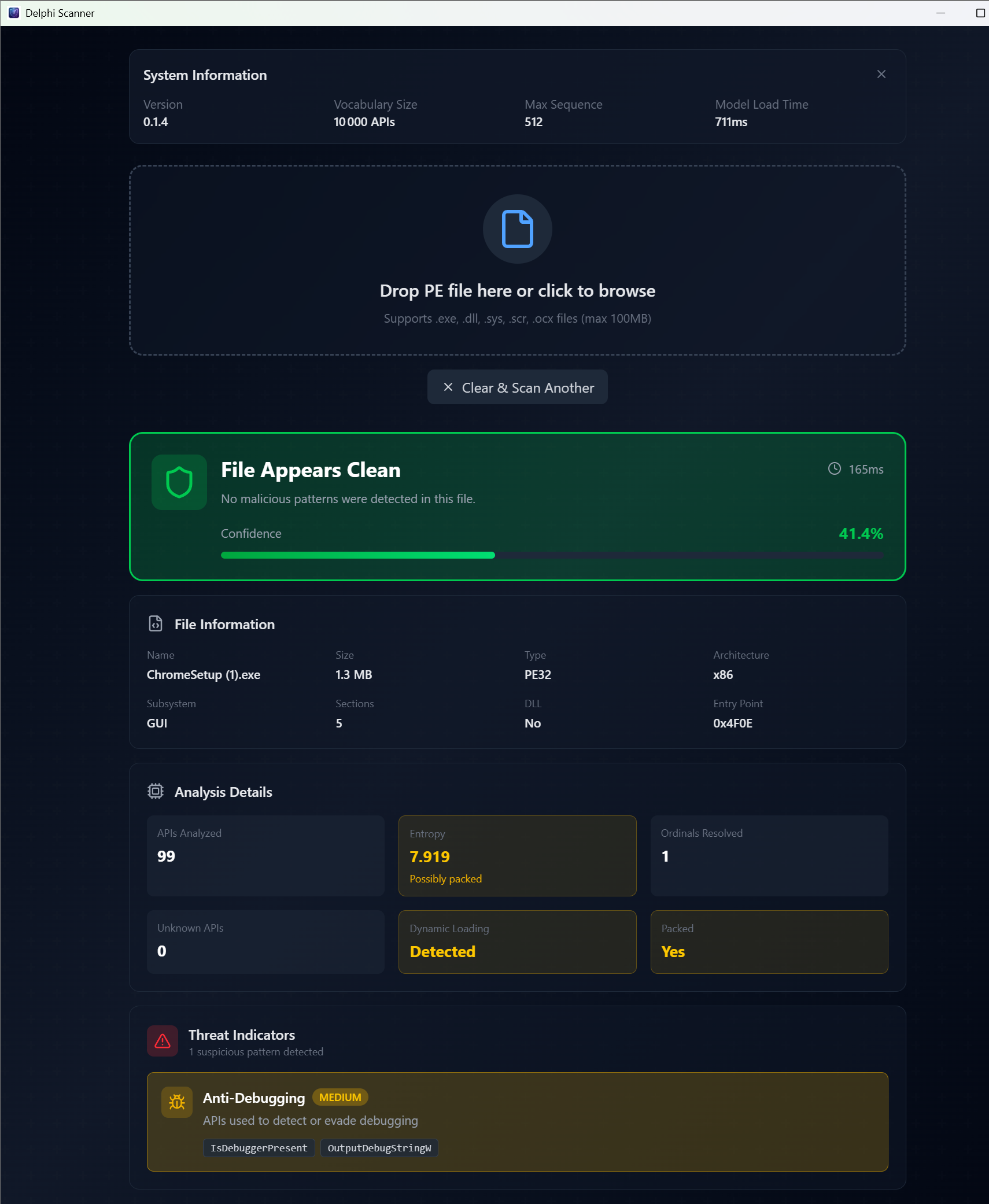}
    \caption{Analysis output for a benign PE file.}
    \label{fig:delphi-benign}
\end{figure}

Figure~\ref{fig:delphi-benign}  shows the analysis results for a legitimate executable (ChromeSetup.exe). The classifier assigns a benign verdict with a relatively low confidence score (41.4\%), reflecting the model’s calibrated uncertainty rather than an overconfident prediction. This behavior is intentional, as benign software may legitimately exhibit characteristics that overlap with malware, such as dynamic API loading or elevated entropy due to compression.

The analysis details reveal several noteworthy static indicators: elevated entropy suggesting possible packing, detection of dynamic loading APIs, and the presence of anti-debugging–related functions. These indicators are surfaced in the Threat Indicators section as a medium-severity warning. Importantly, these behavioral flags do not override the machine learning verdict; instead, they provide contextual information to the analyst. This example illustrates how Delphi Scanner avoids false positives by decoupling behavioral interpretation from classification, while still exposing potentially relevant technical signals.

\subsection{Analysis of a Malicious File}
\begin{figure}[t]
    \centering
    \includegraphics[width=0.9\linewidth]{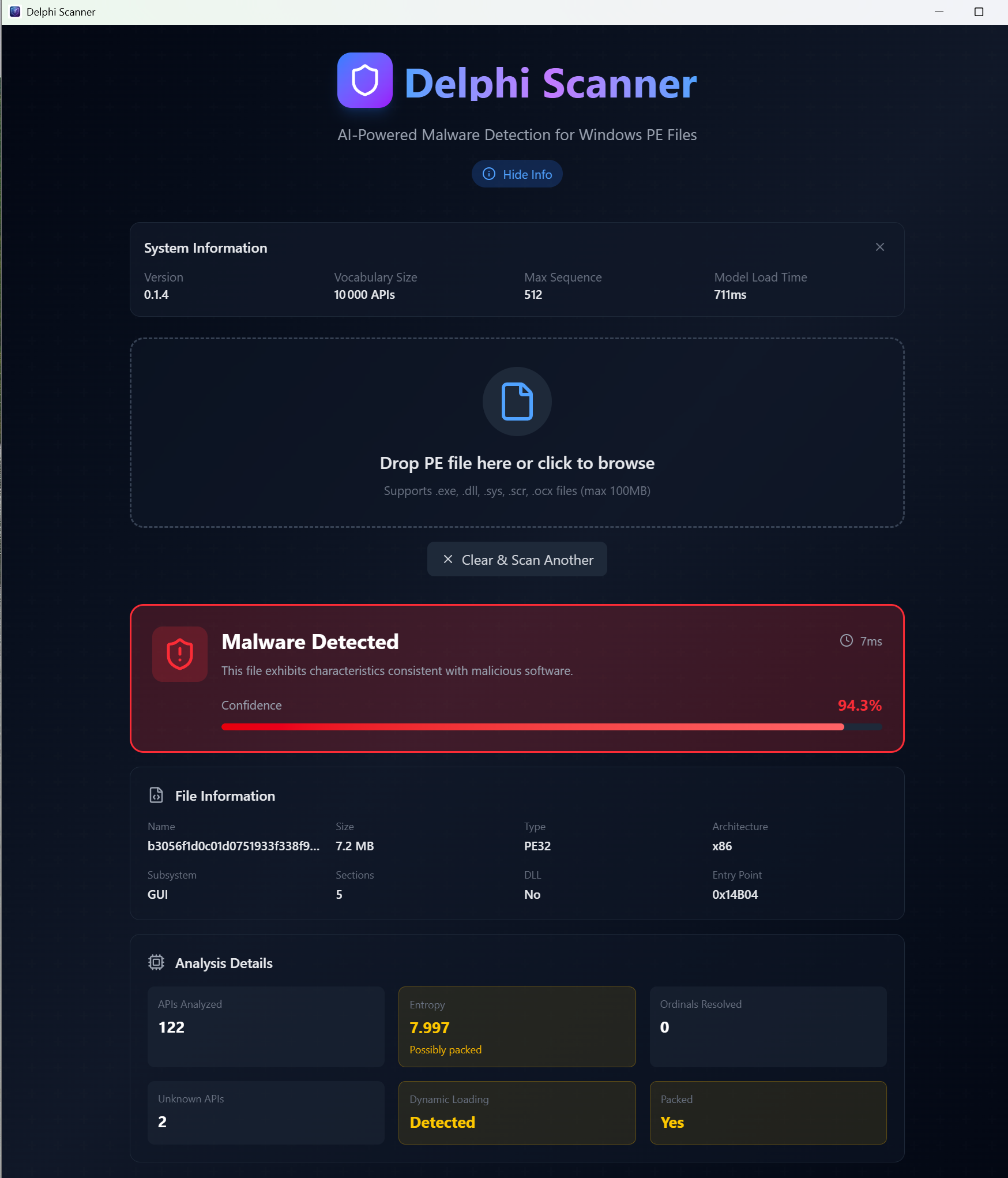}
    \caption{Analysis output for a malicious PE file.}
    \label{fig:delphi-malware}
\end{figure}

Figure~\ref{fig:delphi-malware} illustrates the analysis of a malicious PE file. In this case, the classifier produces a malware verdict with high confidence (94.3\%), indicating strong alignment between the observed API sequence and learned malicious patterns. The static analysis details show a higher number of extracted APIs, elevated entropy consistent with packing, unresolved ordinals, and the presence of unknown APIs, all of which are commonly associated with malicious binaries.

The behavioral interpretation layer reinforces the classification outcome by highlighting suspicious characteristics, including dynamic loading and packing indicators. The combination of a high-confidence verdict and multiple corroborating behavioral signals enables rapid triage and prioritization without requiring execution or dynamic analysis. This example demonstrates how Delphi Scanner supports analyst decision-making by providing both an automated judgment and interpretable evidence.

\subsection{Summary of Case Study Findings}
This case study demonstrates that Delphi Scanner effectively integrates efficient static malware detection with interpretable, analyst-oriented output. Benign files exhibiting suspicious but legitimate characteristics are handled conservatively, while malicious samples are identified with high confidence and supported by clear behavioral indicators. These examples validate the system’s design goals of transparency, low-latency analysis, and practical usability, complementing the quantitative results reported earlier.

\section{Discussion and Limitations}
\label{sec:Discussion and Limitations}  
This section interprets the experimental findings in the context of the system's design objectives, discusses the boundaries of the proposed approach, and outlines directions for future research. 

\subsection{Robustness and Evasion Considerations}
The experimental results demonstrate that the CNN-based classifier achieves the best trade-off among all evaluated architectures, combining the highest accuracy (95.35\%) with the smallest model size (1.53~MB) and the lowest inference latency (0.56~ms). This confirms that local API co-occurrence patterns are sufficiently discriminative for static PE malware detection, and that the additional complexity of transformer or recurrent architectures does not translate into measurable gains in this setting.

The robustness evaluation reveals that the model generalizes meaningfully beyond its training distribution, retains strong detection capability under packing-induced feature degradation, and resists all three evaluated adversarial manipulation strategies. The most practically significant finding is the recall gap between packed and unpacked samples, which reflects the reduced import table visibility caused by packing rather than a fundamental weakness in the learned representations. Resistance to API benign injection is particularly noteworthy, as this attack directly targets the model's feature space yet produces only a modest recall reduction, suggesting that the classifier captures discriminative local patterns that remain robust to benign noise injection.

\subsection{Limitations of Static Analysis}
Static inspection is inherently constrained to metadata present in the binary at analysis time. Packed executables expose only unpacking stub imports, limiting feature coverage, as confirmed empirically by the 5.08 percentage point recall gap observed between packed and unpacked variants. Runtime behaviors such as dynamically resolved APIs, downloaded payloads, and self-modifying code remain unobservable without execution. These constraints are structural to static analysis and represent a deliberate trade-off against the speed, safety, and privacy benefits of local offline analysis. Consequently, malware relying exclusively on dynamic API resolution at runtime remains outside the scope of import-based static analysis and constitutes the most significant open robustness challenge.

\subsection{Dataset Bias and Generalization}
The 89.13\% recall on MalwareBazaar samples confirms meaningful generalization to out-of-distribution threats, while the gap relative to in-distribution performance reflects the expected effect of distributional shift toward newer malware families~\cite{generalization_ids}. Nevertheless, the training dataset remains subject to temporal and family-level biases, and a temporally stratified evaluation training on earlier samples and testing on a later time window, remains an important direction for future work. Addressing longer-term generalization would require continuous dataset updates, broader malware family coverage, and potentially the integration of complementary behavioral signals.

\subsection{Practical Deployment Relative to Existing Solutions}
Delphi Scanner is not intended to replace full-featured antivirus products or cloud-based malware scanning services. Instead, it occupies a complementary operational niche: fast, privacy-preserving, and interpretable pre-execution screening. Its lightweight ONNX backend makes it architecturally extensible to integration within broader security pipelines, including browser-based download screening, antivirus pre-filtering, and endpoint detection and response (EDR) platforms. In a layered defense architecture, it can serve as an efficient first-pass triage component, feeding high-confidence verdicts to downstream dynamic analysis or YARA rule engines. These integration scenarios are enabled by the system's sub-millisecond inference latency and its fully offline, modular design. The inherent trade-offs of this approach — namely the absence of real-time protection, signature-based detection, and continuous threat intelligence updates — reflect a deliberate focus on transparency and deployability rather than comprehensive threat coverage. Beyond standalone desktop deployment, Delphi Scanner's modular architecture supports broader organizational integration scenarios. It can be deployed on internal servers, Network Attached Storage (NAS) systems, or enterprise file-sharing platforms to perform automated pre-execution screening of executable files, limiting the lateral movement of malware within organizational networks. Integration with Cloud Access Security Brokers (CASBs) further extends this capability to monitor executable files shared through cloud-based services, while preserving the system's lightweight footprint and offline inference design.

\subsection{Future Work} 
Several directions emerge naturally from the current limitations. First, the integration of dynamic analysis as a complementary stage would address the most significant gap of the current system: the inability to observe runtime behaviors such as dynamically resolved APIs, downloaded payloads, and self-modifying code. In a layered architecture, Delphi Scanner could serve as a fast pre-screening filter, escalating uncertain or high-entropy samples to a dynamic analysis sandbox, thereby combining the efficiency of static analysis with the behavioral depth of dynamic execution.

Second, integration of lightweight unpacking heuristics would improve detection coverage for packed executables, directly addressing the recall gap observed between packed and unpacked variants in the robustness evaluation.

Third, a temporally stratified evaluation protocol — training on earlier samples and testing on a later time window-would provide rigorous empirical insight into concept drift and real-world generalization to emerging malware families, addressing a limitation acknowledged in the current dataset composition.

Fourth, a structured analyst study measuring triage efficiency with and without the behavioral interpretation layer would provide quantitative validation of its practical utility beyond the qualitative case study presented in Section~\ref{sec:case}.

Finally, future research must address the open challenges associated with dynamic malware detection, including fileless malware that resides entirely in memory, AI-powered malware capable of adapting its behavior to evade detection, and adversarial threats such as data poisoning attacks that gradually degrade model accuracy. Privacy and ethical concerns surrounding the execution of untrusted code further complicate dynamic analysis in practice. Hybrid approaches combining static and dynamic analysis represent a promising direction, though their effective integration remains a technically demanding endeavor.

\section{Conclusion}
\label{sec:conclusion}
This paper presented Delphi Scanner, a static malware detection system for Windows PE files that combines deep learning-based API sequence classification with interpretable behavioral analysis. Experimental results on a large and diverse dataset show that convolutional neural networks provide an effective balance between detection performance and computational efficiency for static API sequence modeling, achieving strong accuracy with a compact model and sub-millisecond inference latency. The decoupled behavioral interpretation layer complements the statistical classifier by translating low-level API evidence into human-readable capability indicators without influencing classification outcomes. Overall, Delphi Scanner demonstrates that static malware detection can be both accurate and interpretable when detection and explanation are explicitly separated, providing a practical foundation for privacy-preserving, analyst-oriented malware triage.\\

\bibliographystyle{splncs04}
\bibliography{references}

\end{document}